\documentclass[11pt,reqno]{amsart}
\usepackage{amsthm,amssymb,amsfonts,dsfont}
\usepackage[left=25mm,top=25mm,bottom=25mm,right=25mm]{geometry}
\usepackage[english]{babel} 
\usepackage[utf8]{inputenc} 
\usepackage{cmap}           
\usepackage{hyperxmp}       
\usepackage{graphicx}
\usepackage{booktabs}
\usepackage{diagbox}
\usepackage{bm}
\usepackage[pdfdisplaydoctitle = true,
      colorlinks = true,
      urlcolor = blue,
      citecolor = blue,
      linkcolor = blue,
      pdfstartview = FitH,
      pdfpagemode = UseNone,
      bookmarksnumbered = true,
      unicode=true]{hyperref} 
\usepackage{tikz}
\usetikzlibrary{shapes} 
\usepackage{color} 
\definecolor{mygreen}{RGB}{28,172,0} 
\definecolor{mylilas}{RGB}{170,55,241}
\definecolor{aliceblue}{rgb}{0.94, 0.97, 1.0}
\definecolor{aquamarine}{rgb}{0.5, 1.0, 0.83}
\definecolor{vert}{RGB}{58, 137, 35}
\definecolor{gris}{RGB}{206, 206, 206} 
\definecolor{rouge}{RGB}{163, 0, 0}
\definecolor{grey}{RGB}{142, 162, 198}
\definecolor{gris_theme}{RGB}{102, 127, 138}
\definecolor{gris_theme_2}{RGB}{88, 110, 120}
\definecolor{bleu_theme}{RGB}{133, 166, 178}
\definecolor{marron}{RGB}{137, 116, 83}
\definecolor{vert}{RGB}{0,100,0}
\definecolor{bleu_pastel}{RGB}{169, 234, 254}
\definecolor{grey}{RGB}{142, 162, 198}
\definecolor{azur_brume}{RGB}{220, 240, 255}
\newtheorem{thm}{Theorem}[section]

\theoremstyle{definition}

\theoremstyle{remark}
\newtheorem{rem}[thm]{Remark}

\numberwithin{equation}{section}

\newcommand{\norm}[1]{\left\Vert#1\right\Vert}
\newcommand{\abs}[1]{\left\vert#1\right\vert}
\newcommand{\set}[1]{\left\{#1\right\}}

\newcommand{\RR}{\mathbb{R}}                            

\newcommand{\OO}{\mathrm{O}}

\newcommand{\octa}{\mathbb{O}}

\renewcommand{\vec}{\pmb}
\newcommand{\mm}{\pmb{m}}
\newcommand{\MM}{\pmb{M}}
\newcommand{\nn}{\pmb{n}}
\newcommand{\uu}{\pmb{u}}

\newcommand{\ee}{\pmb{e}}

\newcommand{\be}{\mathbf{e}}
\newcommand{\balpha}{\pmb{\alpha}}
\newcommand{\bbeta}{\pmb{\beta}}

\newcommand{\bsigma}{\pmb{\sigma}}

\newcommand{\id}{\bm{1}}

\newcommand{\bI}{\mathbf{I}} 
\newcommand{\bJ}{\mathbf{J}} 
\newcommand{\bP}{\mathbf{P}} 

\DeclareMathOperator{\tr}{tr}

\tikzstyle{rect}=[draw=black, fill=white, line width=1.5pt, rectangle,
 rounded corners, inner sep=10pt, inner ysep=5pt]
\tikzstyle{fancytitle}=[fill=black, text=white]

\tikzstyle{diam}=[draw=black, fill=white,rotate=0, line width=1.5pt, diamond, inner sep=10pt, inner ysep=5pt]
\tikzstyle{fancytitle}=[fill=black, text=white]

\tikzstyle{elli}=[draw=black, fill=white,rotate=0, line width=1.5pt, ellipse, inner sep=10pt, inner ysep=5pt]
\tikzstyle{fancytitle}=[fill=black, text=white]

\tikzstyle{rect_algo}=[draw=black, fill=azur_brume, line width=1.5pt, rectangle,
 rounded corners, inner sep=10pt, inner ysep=5pt]

\newcolumntype{L}[1]{>{\raggedright\arraybackslash}m{#1}}
\newcolumntype{C}[1]{>{\centering\arraybackslash}m{#1}}

\hypersetup{
  pdfauthor={J. Taurines, B. Kolev, R. Desmorat, O. Hubert}, 
  pdftitle={Reduced polynomial invariant integrity basis for in-plane magneto-mechanical loading}, 
  pdfsubject={}, 
  pdfkeywords={Invariants, Integrity basis, Strong coupling, Thin structures, Magnetostriction}, 
  pdflang=en, 
  }

\begin{document}

\title[Reduced integrity basis for in-plane magnetostriction]{Reduced polynomial invariant integrity basis \\ for in-plane magneto-mechanical loading}

\author{J. Taurines}
\address[Julien Taurines]{Université Paris-Saclay, ENS Paris-Saclay, CentraleSupélec, CNRS, LMPS - Laboratoire de Mécanique Paris-Saclay, 91190, Gif-sur-Yvette, France}
\email{julien.taurines@ens-paris-saclay.fr}

\author{B. Kolev}
\address[Boris Kolev]{Université Paris-Saclay, ENS Paris-Saclay, CentraleSupélec, CNRS, LMPS - Laboratoire de Mécanique Paris-Saclay, 91190, Gif-sur-Yvette, France}
\email{boris.kolev@ens-paris-saclay.fr}

\author{R. Desmorat}
\address[Rodrigue Desmorat]{Université Paris-Saclay, ENS Paris-Saclay, CentraleSupélec, CNRS, LMPS - Laboratoire de Mécanique Paris-Saclay, 91190, Gif-sur-Yvette, France}
\email{rodrigue.desmorat@ens-paris-saclay.fr}

\author{O. Hubert}
\address[Olivier Hubert]{Université Paris-Saclay, ENS Paris-Saclay, CentraleSupélec, CNRS,  LMPS - Laboratoire de Mécanique Paris-Saclay, 91190, Gif-sur-Yvette, France}
\email{olivier.hubert@ens-paris-saclay.fr}


\date{January 10, 2022}%


\begin{abstract}
  The description of the behavior of a material subjected to multi-physics loadings requires the formulation of constitutive laws that usually derive from Gibbs free energies, using invariant quantities depending on the considered physics and material symmetries. On the other hand, most of crystalline materials can be described by their crystalline texture and the associated preferred directions of strong crystalline symmetry (the so-called \textit{fibers}). Moreover, among the materials produced industrially, many are  manufactured in the form of sheets or of thin layers. This article has for object the study of the magneto-mechanical coupling which is a function of the stress $\bsigma$ and the magnetization $\MM$. We consider a material with cubic symmetry whose texture can be described by one of three fibers denoted as $\theta$, $\gamma$ or $\alpha'$, and which is thin enough so that both the stress and the magnetization can be considered as in-plane quantities. We propose an algorithm able to derive linear relations between the 30 cubic invariants $I_{k}$ of a minimal integrity basis describing a magneto-elastic problem, when they are restricted to in-plane loading conditions and for different fiber orientations. The algorithm/program output is a reduced list of invariants of cardinal 7 for the \{100\}-oriented $\theta$ fiber, of cardinal 15 for the \{110\}-oriented $\alpha'$ fiber and of cardinal 8 for the \{111\}-oriented $\gamma$ fiber. This reduction (compared to initial cardinal 30) can  be of great help for the formulation of low-parameter macroscopic magneto-mechanical models.
\end{abstract}

\maketitle

\section{Introduction}

Since the discovery of the crystalline nature of metals, and of the anisotropic nature of the associated behaviors, metallurgists have sought to improve thermomechanical treatments in order to develop the most favorable crystallographic textures. Research is carried out, in particular, on magnetic materials~\cite{chikazumi1997,HS2008}. In this regard, we can cite the well-known \textit{Goss} texture for 3\%silicon-iron alloys~\cite{Lit1982} used as vehicles of the magnetic flux in high power transformers: the magnetic permeability in the rolling direction is greatly improved comparing to the magnetic permeability of an isotropic 3\%silicon-iron polycrystal; the coercive field is considerably reduced, leading to a drastic decrease of energy losses per magnetization cycle. References~\cite{JXZ2020,TC2021,Qiao2021} thus report recent developments in this field. On the other hand, cold rolled and annealed FeNi alloys are known to be able to develop a so-called \textit{cube} texture during recrystallization~\cite{ABHB2014,Wang2020}, meaning that the crystallographic frame coincides with the sheet frame: this texture leads to a high magnetic permeability in both the rolling and transversal (to the rolling) directions. For their part, magnetic shape memory alloys (MSMA) are generally produced as single crystalline bulk materials since polycrystals exhibit lower magneto-mechanical and fatigue performances. Recent works show the possibility of producing hypertextured polycrystalline Ni-Mn-Ga MSMA and possibly in the form of thin layers, opening a wide new range of applications~\cite{Li2014}. Conversely, the crystallographic texture can be high and uncontrolled. This is frequently observed for very thin magnetic materials used in high frequency electronic systems (the small thickness allows for a better homogeneity of the electromagnetic fields through the thickness at very high frequency, typically GHz). The textures encountered may vary but generally follow epitaxy rules (depending on the sublayer orientation): the direction normal to the layer is frequently a direction of strong crystalline symmetry~\cite{Cates1994,YKI2006,HMTN2008}. Finally, the increasing miniaturization of electronic systems is pushing for the use of small size --therefore thin-- magnetic components. We thus observe an increasing scale confusion between component and crystals, making behaviors very sensitive to surface effects~\cite{HS2008,Hub2008}.

Moreover, just like a magnetic material magnetizes under the effect of a magnetic field, it deforms. This deformation, called magnetostriction strain, is the first manifestation of the magneto-elastic coupling~\cite{DTdLac1993}. The inverse magneto-elastic coupling is the effect of a mechanical stress on the magnetic behavior (the Villari effect)~\cite{Boz1951,Cul1972}. Some materials such as 3\%silicon-iron or 27\%cobalt-iron alloys develop such a sensitivity to mechanical loading that a second-order phenomenon appears (the morphic effect~\cite{dTdLac1982}):  increasing magnetic permeability with increasing stress at low stress level, then decreasing magnetic permeability with increasing stress at higher stress level. The introduction of a second-order (quadratic both in $\bsigma$ and in $\MM$) magneto-elastic coupling term in the expression of the Gibbs free energy density makes it possible to account for this effect: simulations and model-experiment comparisons have been proposed in~\cite{Hub2019} using a second-order isotropic approximation. The isotropic approximation is however reductive given the cubic symmetry of the medium. The development of a cubic second order term however requires the use of a 6th order tensor. Its construction and handling are difficult. The approach by \emph{Invariant Theory} and the use of a \emph{minimal} integrity basis (see \cite{OKA2017,DA2019}) are the core of a recent article dealing with magneto-elastic coupling in cubic media~\cite{TOD2021}. It allows for a rigorous construction of Gibbs free energy density at any order without missing any term.

However, this integrity basis has a large cardinal (= 30), which can make the identification process very cumbersome, when higher order terms are involved. In this paper, we consider thin textured sheets for which the integrity basis given in \cite{TOD2021} can be reduced when only in-plane magneto-mechanical loadings are considered. Indeed, the magnetization of the material is assumed to remain in the sheet plane due to the strong demagnetizing fields created by any emergent magnetization~\cite{HS2008}. This boundary condition is completed by the usual plane stress assumption.

The paper is organized as follows. In \autoref{sec:textures}, we present the main concepts related to the definition of a crystallographic texture for materials with cubic symmetry. In \autoref{sec:magneto-elasticity}, we introduce cubic magneto-elasticity energy densities and the fundamental 30 invariants $I_{k}$ obtained in~\cite{TOD2021} which are necessary to formulate them. The mathematical formalism used to describe in-plane loadings and reduce the number of fundamental invariants for these loadings is introduced in \autoref{sec:in-plane-loadings}. The results for the three main material fibers are provided in details in \autoref{sec:reduced-generating-sets}. Finally, an algorithm and its implementation in \emph{Macaulay2} to obtain relations between the evaluated invariants $I_{k}$ for some given crystallographic textures is proposed in \autoref{sec:algo}. The output is a minimal list of polynomial invariants that allows for the most general expression of Gibbs free energy density to be formulated for a large set of crystallographic textures.

\section{Texture and orientation data function of crystalline materials}
\label{sec:textures}

\begin{figure}[ht]
  \centering
  \includegraphics[width=0.4\textwidth]{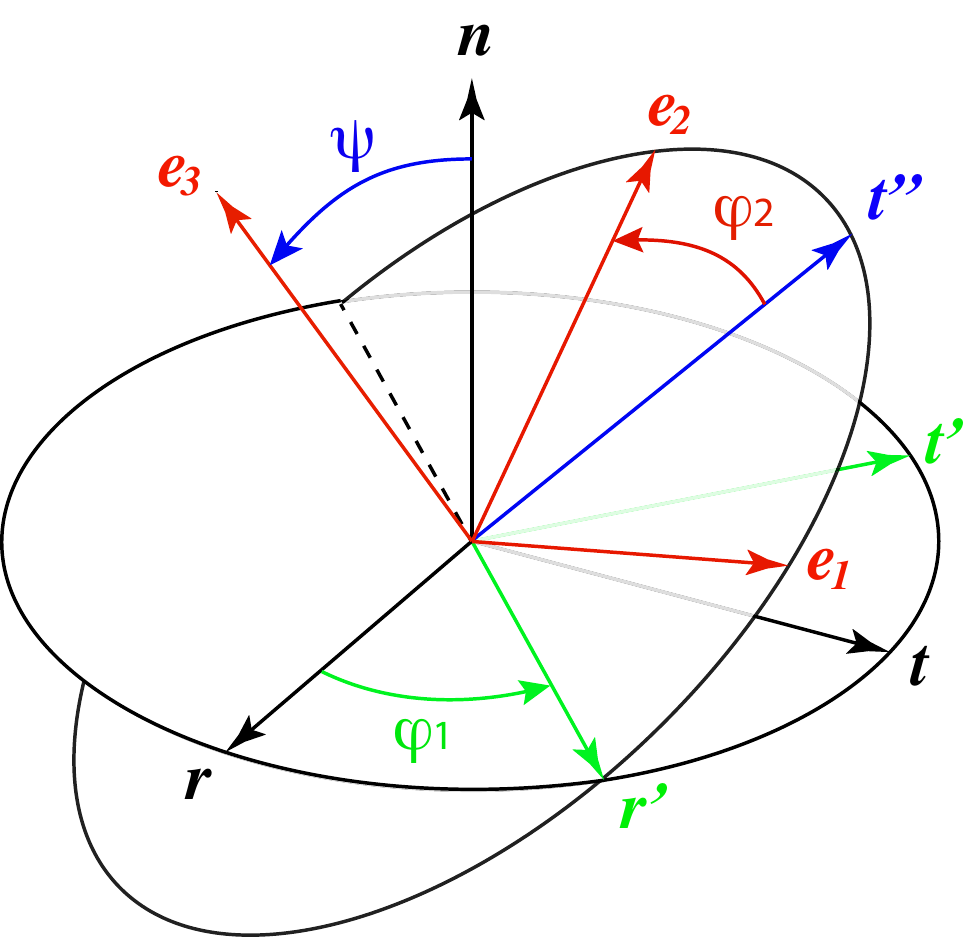}
  \caption{Illustration of reference $(\vec r, \vec t, \vec n)$=(\textit{RD},\textit{TD},\textit{ND}) and rotated $(\ee_{1}, \ee_2, \ee_3)$ frames using ($\varphi_{1},\psi,\varphi_2$) Euler angles and Bunge rotation rules.}
  \label{fig:euler}
\end{figure}

The crystallographic texture is a simplified description of how the individual crystallites that make up the material are distributed.  In materials science, Euler angles are used to describe a single crystal orientation relative to the axes of the sample (as the reference orthonormal frame). The following denomination is usually employed and illustrated in \autoref{fig:euler}: $\vec r=$\textit{RD} for Rolling Direction, $\vec t=$\textit{TD} for Transversal Direction and $\vec n=$\textit{ND} for Normal Direction indicate the reference orthonormal frame. Such denominations are obviously borrowed from sheets metallurgy and rolling process.

The three angles defining the orientation of crystal axes relative to the reference frame are noted (${\displaystyle \varphi_{1}, \psi, \varphi_{2}}$) using Bunge representation:
\begin{itemize}
  \item $\varphi_{1}$ corresponds to a first rotation operation around \textit{ND} axis; the new coordinate system is named $(\vec r',\vec t',\nn)$.
  \item $\psi$ corresponds to a second rotation operation around $\vec r'$ axis; the new coordinate system is named $(\vec r',\vec t'',\ee_3)$.
  \item $\varphi_{2}$ corresponds to a third and last rotation operation around $\ee_3$ axis; the new frame is denoted $(\ee_{1}, \ee_2, \ee_3)$.
\end{itemize}
Associating the cubic crystallographic frame ([100] [010] [001]) to the orthonormal frame $(\ee_{1}, \ee_2, \ee_3)$, it is possible to observe some connection between a crystallographic axis $[uvw]$ and the principal axes of the reference frame. We thus usually designate by the following Miller indices combination
\begin{equation*}
  \{hkl\} <uvw>
\end{equation*}
a situation where $\{hkl\}$ corresponds to \textit{ND} (for cubic symmetry, $h$, $k$ and $l$ also indicate the components of the vector normal to the sheet plane) and where $<uvw>$ corresponds to \textit{RD}.

\begin{figure}[ht]
  \centering
  \includegraphics[width=0.6\textwidth]{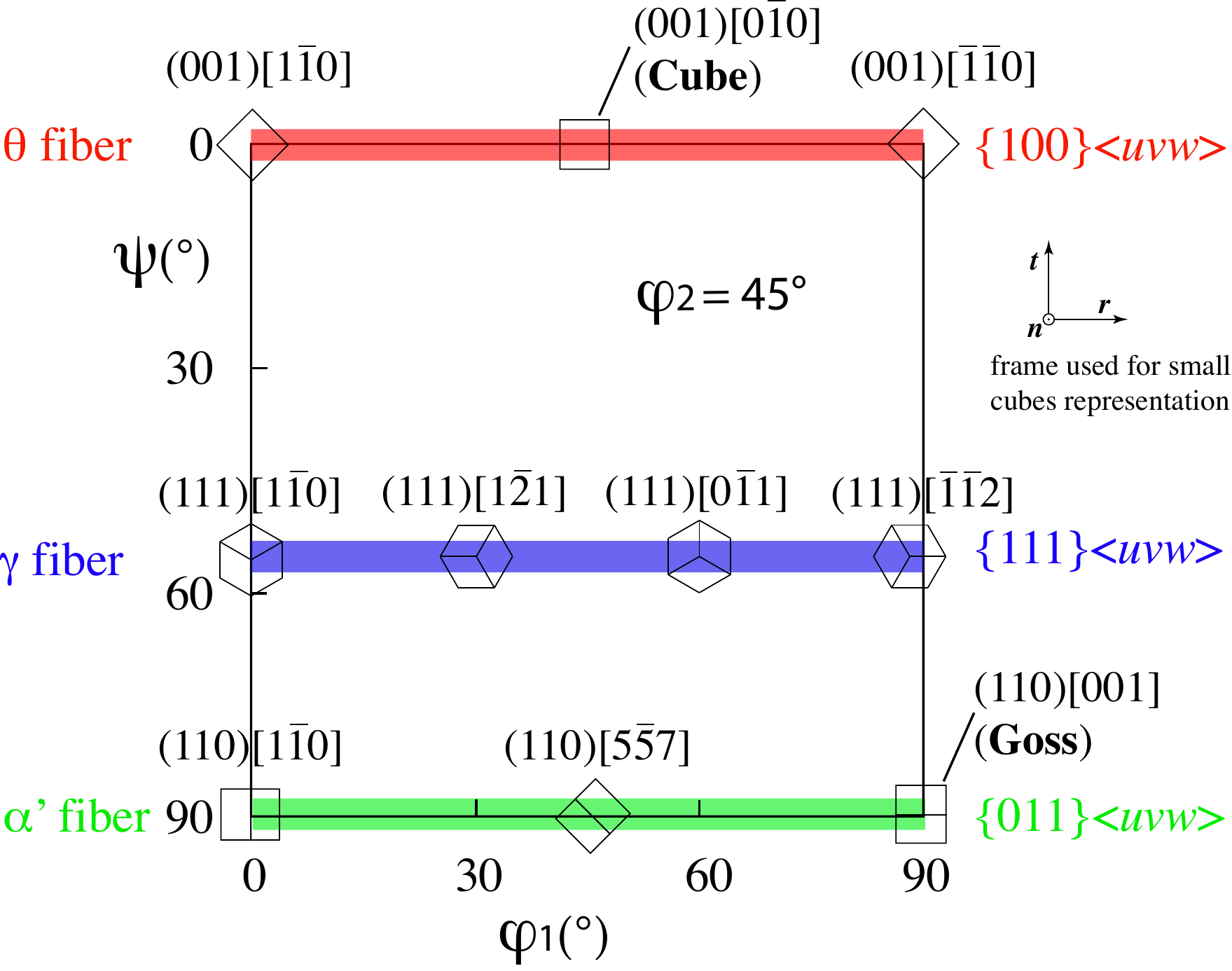}
  \caption{Cut of Euler space for $\varphi_2 = 45^\circ$ and highlighting of some major texture components.}
  \label{fig:IPF}
\end{figure}

Some remarkable directions do obviously correspond to a given set of Euler angles. It is thus possible to place some of these remarkable directions in a ($\varphi_{1},\psi$) plane for a fixed $\varphi_2$. \autoref{fig:IPF} illustrates a cut of the Euler space for $\varphi_2 = 45^\circ$, highlighting some remarkable orientations. The three main situations colored in red, green and blue are:

\begin{itemize}
  \item \textcolor{red}{$\theta$ fiber}: \{100\}$<uvw>$ - the normal plane corresponding to a cube face;
  \item \textcolor{green}{$\alpha'$ fiber}: \{110\}$<uvw>$ - the normal plane corresponding to a cube diagonal plane;
  \item \textcolor{blue}{$\gamma$ fiber}: \{111\}$<uvw>$ - the normal direction corresponding to a cube trisectrix.
\end{itemize}

When specific planes and directions are considered (and not direction and plane families), parentheses and brackets are used. We highlight the following \textit{classical} textures:

\begin{itemize}
  \item Cube texture, belonging to $\theta$ fiber: (001)$[0\bar{1}0]$;
  \item Goss texture, belonging to $\alpha'$ fiber: (110)$[001]$.
\end{itemize}

Of course, most materials do not present a single \{hkl\} $<uvw>$ orientation but a set of orientations, corresponding to a set of Euler angles and defining the orientation data function (ODF). A material is denoted as \textit{textured} when this distribution is tightened on one or more particular directions. Otherwise, it is a non-textured material. Magnetic materials obtained by rolling process or deposit thus often present remarkable textures that can be found in the above list \footnote{Other textures are referenced in literature -so called \textit{brass} texture, \textit{copper} texture, $\alpha$ fiber, $\epsilon$ fiber and so on -  but they do not exhibit a single crystallographic direction perpendicular to the sheet plane.}. Some others may be described by a combination of these textures~\cite{NABI2014,DBH2020}. A texture can thus represent, beyond a simple assembly of crystals, the material itself, thereby allowing for a macroscopic modeling, ignoring its multiscale nature. Within the framework of the development of magnetoelastic constitutive laws of any order, taking into account the existence of a texture can lead to significant simplifications. We focus our efforts on the three main fibers listed above.

\section{Cubic magneto-elasticity}
\label{sec:magneto-elasticity}

There are two main manifestations of magnetoelastic coupling in magnetic materials: the magnetostriction strain and the variation of the magnetization  under stress. The modeling which is proposed in the literature is generally based either on constitutive tensors (of order $3$ and higher~\cite{Hub2019,ME2021}) or on well chosen invariants~\cite{Tou1956,FM1996,FBK2010,DOS2004,DKT2012,BSR2021}. Such a modeling can also be carried out at different scales from microscopic to macroscopic. At the crystal scale, magnetization is associated with the cubic (octahedral) symmetry group $\octa$ which is defined by
\begin{equation*}
  \octa = \set{g\in \OO(3);\; g \ee_{i} = \pm \ee_{j}},
\end{equation*}
where $(\ee_{i})$ is the canonical orthonormal basis of $\RR^{3}$ and $\OO(3)$ is the orthogonal group. This group is of order $48$ : it contains $24$ rotations and $24$ orientation-reversing isometries leaving the cube invariant~\cite{KS1974,KE1990,WG2004,TOD2021}. The stress tensor $\bsigma$ and the magnetization pseudo-vector $\mm$ (with $\norm{\mm}=m_s$ the saturation magnetization) are almost homogeneous at this scale. The macroscopic behavior, \emph{i.e.}, at the representative volume element (RVE) scale of volume $V$, is obtained by an homogenization process~\cite{DHBB2008,DHW2014,Hub2019}. The macroscopic magnetization is, for example, given by
\begin{equation}
  \MM = \langle \mm \rangle=\frac{1}{V}\int_V\mm\,dv.
\end{equation}
A direct description of the macroscopic behavior by using a well chosen expression of the Gibbs free energy density is an alternative approach. For instance, isotropic energy densities have been proposed in~\cite{Sab1993,Bela2017,BSR2021}, transversely isotropic ones  in~\cite{DKT2011} and orthotropic ones in~\cite{QLZ2020}. Cubic invariants of the pair $(\mm,\bsigma)$~\cite{SSR1963,TOD2021}, fully relevant at the magnetic domains scale, may also be relevant at the macroscopic scale, by considering the macroscopic magnetization $\MM$ in place of the local magnetization $\mm$, \emph{if cubic symmetry applies at the macroscopic scale}. As extensively explained in the introduction of this paper, this situation is encountered in single crystals, or when the material, consisting of an assembly of cubic crystals, is highly textured.

Classical Invariant Theory~\cite{Stu1993,Olver1999,DK2015} is a robust and efficient tool which helps to formulate Gibbs energy densities that respect the material symmetry, the key points of this theory being
\begin{enumerate}
  \item the choice of the relevant group $G$  for the material symmetry/physics~\cite{LL1960,Bir1964,SC1984,WG2004},
  \item the determination of an integrity basis $\set{I_{k}}$ or, more generally, of a functional basis~\cite{Boe1987,Wey1997}.
\end{enumerate}
Therefore, each magneto-elasticity energy density which is $G$-invariant can be expressed as
\begin{equation}\label{Gibbsinv}
  \Psi^{\mu\sigma}=\Psi^{\mu\sigma}(I_{k}).
\end{equation}

In the following, we consider a cubic microstructure, \emph{i.e.}, a microstructure which is invariant by the octahedral symmetry group $G=\octa$, and where the set $\set{I_{k}}$ is the integrity basis for polynomial cubic invariants in $\MM$ and $\bsigma$~\cite{SSR1963,KS1974,KE1990} provided in \autoref{tab:even-inv} (see also~\cite{WG2004,TOD2021} for how accounting for magnetic point groups~\cite{SC1984,Mau1990}). Transversely isotropic (resp. orthotropic) magneto-elastic energy densities are handled in the same way, but by considering transversely isotropic (resp. orthotropic) invariants instead of cubic ones (see~\cite{Adk1959,Adk1960,DKT2012}).

\begin{table}[ht]
  \centering
  \begin{tabular}{cccc}
    \toprule
    deg($\MM$) & deg($\bsigma$) & Formula                                                                                                                           & Tri-graded notation
    \\
    \midrule
    0          & 1              & \footnotesize $\tr\bsigma$                                                                                                        & --
    \\
    0          & 2              & \footnotesize $\bsigma^{\bar d} : \bsigma^{\bar d}$                                                                               & $I_{002}$
    \\
    0          & 2              & \footnotesize $\bsigma^{d}: \bsigma^{d}$                                                                                          & $I_{020}$
    \\
    0          & 3              & \footnotesize $\tr(\bsigma^{\bar d\, 3})$                                                                                         & $I_{003}$
    \\
    0          & 3              & \footnotesize $\bsigma^{\bar d\, 2} :\bsigma^{d}$                                                                                 & $I_{012}$
    \\
    0          & 3              & \footnotesize $\tr(\bsigma^{d\,3})$                                                                                               & $I_{030}$
    \\
    0          & 4              & \footnotesize $ (\bsigma^{\bar d\, 2})^{\bar d} : ( \bsigma^{\bar d\, 2})^{\bar d}$
               & $I_{004}$
    \\
    0          & 4              & \footnotesize $ \tr( \bsigma^{\bar d}  \bsigma^{d} \bsigma^{\bar d}\bsigma^{d} )$                                                 & $I_{022}$
    \\
    0          & 5              & \footnotesize $\big( \bsigma^{\bar d} (\bsigma^{\bar d\, 2})^{\bar d} \bsigma^{\bar d}\big):\bsigma^{d}$                          & $I_{014}$
    \\
    2          & 0              & \footnotesize $ \Vert \MM \Vert^{2}$                                                                                              & $I_{200}$
    \\
    2          & 1              & \footnotesize $(\MM \otimes \MM)^{\bar d}: \bsigma^{\bar d}$                                                                      & $I_{201}$
    \\
    2          & 1              & \footnotesize $ (\MM \otimes \MM)^d : \bsigma^{d} $                                                                               & $I_{210}$
    \\
    2          & 2              & \footnotesize $(\MM \otimes \MM)^{d}: \bsigma^{\bar d\, 2} $                                                                      & $I_{202}^{a}$
    \\
    2          & 2              & \footnotesize $ (\MM \otimes \MM)^{\bar d}:\bsigma^{\bar d\, 2} $                                                                 & $I_{202}^{b}$
    \\
    2          & 2              & \footnotesize $(\MM \otimes \MM)^{\bar d} :(\bsigma^{\bar d}\bsigma^{d})$                                                         & $I_{211}$
    \\
    2          & 2              & \footnotesize $(\MM \otimes \MM)^d: \bsigma^{d\, 2}$                                                                              & $I_{220}$
    \\
    2          & 3              & \footnotesize $(\MM \otimes \MM)^{\bar d}:\big((\bsigma^{\bar d\, 2})^{\bar d}\bsigma^{\bar d} \big)$                             & $I_{203}$
    \\
    2          & 3              & \footnotesize $(\MM \otimes \MM)^d:\big((\bsigma^{\bar d\, 2})^{d}\bsigma^{d} \big)$                                              & $I_{212}^{a}$
    \\
    2          & 3              & \footnotesize $(\MM \otimes \MM)^{\bar d}:\big((\bsigma^{\bar d\, 2})^{\bar d}\bsigma^{d} \big)$                                  & $I_{212}^{b}$
    \\
    2          & 3              & \footnotesize $(\MM \otimes \MM)^{\bar d}:\big(\bsigma^{d}\bsigma^{\bar d}\bsigma^{d}\big)$                                       & $I_{221}$
    \\
    2          & 4              & \footnotesize $(\MM \otimes \MM)^{d}:\big(\bsigma^{\bar d}(\bsigma^{\bar d\, 2})^{\bar d}\bsigma^{\bar d}\big)$                   & $I_{204}$
    \\
    2          & 4              & \footnotesize $(\MM \otimes \MM)^{\bar d}: \big((\bsigma^{\bar d\, 2})^{d} \bsigma^{\bar d }\bsigma^{d} \big)$                    & $I_{213}$
    \\
    2          & 4              & \footnotesize $(\MM\otimes \MM)^{\overline{d}} :\big( \bsigma^{d} (\bsigma^{\overline{d}\, 2})^{\overline{d}} \bsigma^{d}\big)  $ & $I_{222}$
    \\
    4          & 0              & \footnotesize $ (\MM \otimes \MM)^{\bar d}:(\MM \otimes \MM)^{\bar d}$                                                            & $I_{400}$
    \\
    4          & 1              & \footnotesize $ (\MM \otimes \MM)^{\bar d\, 2}:\bsigma^{\bar d} $                                                                 & $I_{401}$
    \\
    4          & 1              & \footnotesize $(\MM \otimes \MM)^{\bar d\, 2}:\bsigma^{d} $                                                                       & $I_{410}$
    \\
    4          & 2              & \footnotesize $(\MM \otimes \MM)^{\bar d\, 2}:\big(\bsigma^{\bar d\, 2}\big)^{\bar d}$                                            & $I_{402}$
    \\
    4          & 2              & \footnotesize $(\MM \otimes \MM)^{\bar d\, 2}:\big(\bsigma^{d}\bsigma^{\bar d}\big)$                                              & $I_{411}$
    \\
    6          & 0              & \footnotesize $ \tr\big( (\MM \otimes \MM)^{\bar d\, 3} \big)$                                                                    & $I_{600}$
    \\
    6          & 1              & \footnotesize $\tr \big( (\MM \otimes \MM)^{d}(\MM \otimes \MM)^{\bar d} (\MM \otimes \MM)^{d}\bsigma^{\bar d}  \big)$            & $I_{601}$
    \\
    \bottomrule
  \end{tabular}
  \caption{A minimal integrity basis~\cite{TOD2021} of $\octa$-invariants for $(\MM,\bsigma)$.}
  \label{tab:even-inv}
\end{table}

A minimal integrity basis of 30 polynomials for cubic invariant polynomials in the pair $(\MM, \bsigma)$ has been obtained first by Smith, Smith and Rivlin in~\cite{SSR1963}, and expressed in the components $M_{i}$ and $\sigma_{ij}$ of $\MM$ and $\bsigma$. Recently~\cite[Theorem 2.10 and Remark 2.11]{TOD2021}, we have proposed an alternative minimal integrity basis of invariants expressed using intrinsic tensorial expressions, rather than components. They are provided in \autoref{tab:even-inv}. To obtain these invariants, the following decomposition of the stress tensor, which was introduced in~\cite{Bertram96}
\begin{equation*}
  \bsigma =  \bsigma^{d} + \bsigma^{\overline{d}} +\frac{1}{3} (\tr\bsigma) \id,
\end{equation*}
has been used. In the canonical cubic basis $(\ee_{1},\ee_{2},\ee_{3})$,
\begin{equation*}
  \bsigma^{\overline{d}}=
  \begin{pmatrix}
    0           & \sigma_{12} & \sigma_{13} \\
    \sigma_{12} & 0           & \sigma_{23} \\
    \sigma_{13} & \sigma_{23} & 0
  \end{pmatrix},
  \qquad
  \bsigma^{d}=
  \begin{pmatrix}
    \sigma_{11}' & 0            & 0            \\
    0            & \sigma_{22}' & 0            \\
    0            & 0            & \sigma_{33}'
  \end{pmatrix}.
\end{equation*}
This decomposition is stable under the action of $\octa$. In particular, for cubic materials, the deviatoric stress tensor $\bsigma' = \bsigma - \frac{1}{3} (\tr\bsigma) \id $ splits into
\begin{equation*}
  \bsigma' = \bsigma^{d} + \bsigma^{\overline{d}}.
\end{equation*}

\begin{rem}
  The introduction of the following two fourth order tensors $\bP_\octa^{\overline{d}}$ and $ \bP_\octa^{d}=\bJ- \bP_\octa^{\overline{d}}$, where
  \begin{equation*}
    \bP_\octa^{\overline{d}} := \frac{1}{2} \sum_{i< j}\be_{ij}\otimes \be_{ij},
    \qquad
    \be_{ij} :=\ee_{i}\otimes\ee_{j}+\ee_{j}\otimes \ee_{i} \; (i\neq j).
  \end{equation*}
  and $\bJ=\bI-\frac{1}{3} \id \otimes \id$ is the deviatoric projector, removes the dependency of this decomposition to the canonical basis $(\ee_{i})$. Indeed, $\bP_\octa^{\overline{d}}$ and $ \bP_\octa^{d}$ correspond to the two orthogonal projectors of $\bsigma$ onto the $\octa$-irreducible components $\bsigma^{\overline{d}}$ and $\bsigma^{d}$~\cite{Ryc1984,Fra1995,DM2011} in any orthonormal frame:
  \begin{equation*}
    \begin{cases}
      \bsigma^{d} := \bP_{\octa}^{d} : \bsigma
      \\
      \bsigma^{\overline{d}} :=\bP_{\octa}^{\overline{d}}:\bsigma
    \end{cases}
  \end{equation*}
  This result is axis $(\ee_{i})$ independent.
\end{rem}

\section{In-plane stress and magnetization for plates with strong crystallographic textures}
\label{sec:in-plane-loadings}

General three-dimensional magneto-elasticity laws have been proposed in~\cite{TOD2021} for alloys composed of cubic symmetry crystals. Due to the quite high ($=30$) cardinal of the minimal integrity basis $\set{I_{k}}$, recalled in \autoref{tab:even-inv}, a large number of material parameters --- quantified for polynomial energy densities --- has been introduced.

\begin{table}[ht]
  \centering
  \begin{tabular}{c|ccccccccccc}
    \toprule
    \diagbox{$\textrm{deg}(\MM)$}{$\textrm{deg}(\bsigma)$} & 0  & 1 & 2  & 3  & 4   & 5   & 6   & 7   & 8    & 9    & 10
    \\
    \midrule
    0                                                      & -- & 1 & 3  & 6  & 11  & 18  & 32  & 48  & 75   & 111  & 160
    \\
    2                                                      & 0  & 2 & 6  & 14 & 31  & 60  & 106 & 180 & 288  & 442  & 659
    \\
    4                                                      & 1  & 3 & 10 & 24 & 53  & 102 & 185 & 312 & 504  & 777  & 1161
    \\
    6                                                      & 1  & 4 & 13 & 34 & 73  & 144 & 262 & 444 & 717  & 1112 & 1660
    \\
    8                                                      & 1  & 5 & 17 & 42 & 95  & 186 & 378 & 576 & 933  & 1443 & 2162
    \\
    10                                                     & 1  & 6 & 20 & 52 & 115 & 228 & 375 & 708 & 1146 & 1748 & 2661
    \\
    \bottomrule
  \end{tabular}
  \caption{Number of material parameters~\cite{TOD2021} for polynomial energy densities with given bi-degree in $(\MM, \bsigma)$.}
  \label{tab:coeffs:mat}
\end{table}

When studying thin sheets or layers and considering in-plane stress and magnetization, some relations may appear between the evaluated invariants, leading to a redundancy of information in the Gibbs energy density $\Psi(I_{k})$. A beforehand rewriting of $\Psi$, involving only the invariants that cannot be rewritten when evaluated for the considered in-plane loading, as functions of some other $I_{k}$ is then necessary for an efficient (low-parameter) modeling. These relations depend on the crystallographic texture of the sheet, \emph{i.e.}, on the considered fiber ($\theta$, $\alpha'$ or $\gamma$, see \autoref{fig:IPF}).

It is obvious that the form of the stress $\bsigma$ and of the magnetization $\MM$ that  satisfy the in-plane conditions are
\begin{equation}
  \label{eq:in-plane}
  \bsigma\cdot  \nn =0
  \quad \textrm{and} \quad
  \MM\cdot  \nn =0.
\end{equation}
We note stress and magnetization: $\bsigma_\theta$, $\MM_\theta$ (for $\nn=\nn_\theta$),
$\bsigma_{\alpha'}$, $\MM_{\alpha'}$ (for $\nn=\nn_{\alpha'}$) and $\bsigma_\gamma$, $\MM_\gamma$
(for $\nn=\nn_\gamma$),
for the three fibers $\theta$, $\alpha'$ and $\gamma$. More precisely, the following equations describe the in-plane hypothesis \eqref{eq:in-plane} for these three cases,
\begin{align}
  \label{eq:plane:100}
   & \bsigma_\theta=
  \begin{pmatrix}
    \sigma_{11} & \sigma_{12} & 0 \\
    \sigma_{12} & \sigma_{22} & 0 \\
    0           & 0           & 0
  \end{pmatrix} ,
   &                    &
  \MM_\theta=
  \begin{pmatrix}
    M_{1} \\
    M_2   \\
    0
  \end{pmatrix} ,
  \\
  \label{eq:plane:110}
   & \bsigma_{\alpha'}=
  \begin{pmatrix}
    \sigma_{11}  & \sigma_{12}  & -\sigma_{12} \\
    \sigma_{12}  & -\sigma_{23} & \sigma_{23}  \\
    -\sigma_{12} & \sigma_{23}  & -\sigma_{23}
  \end{pmatrix} ,
   &                    &
  \MM_{\alpha'}=
  \begin{pmatrix}
    M_{1} \\
    M_2   \\
    -M_2
  \end{pmatrix} ,
  \\
  \label{eq:plane:111}
   & \bsigma_\gamma=
  \begin{pmatrix}
    -\sigma_{12}-\sigma_{13} & \sigma_{12}              & \sigma_{13}              \\
    \sigma_{12}              & -\sigma_{12}-\sigma_{23} & \sigma_{23}              \\
    \sigma_{13}              & \sigma_{23}              & -\sigma_{13}-\sigma_{23}
  \end{pmatrix},
   &                    &
  \MM_\gamma=
  \begin{pmatrix}
    M_{1} \\
    M_2   \\
    -M_{1}-M_2
  \end{pmatrix} .
\end{align}
Expressions are given in the canonical cubic symmetry basis $(\ee_{1},\ee_2,\ee_3)$ of \autoref{fig:euler}, with $\nn_\theta=\ee_3$, $\nn_{\alpha'}=\frac{1}{\sqrt{2}}(\ee_2+\ee_3)$ and $\nn_\gamma=\frac{1}{\sqrt{3}}(\ee_{1}+\ee_2+\ee_3)$.

The assumption of plane stress leads to the reduction of the number of independent components $\sigma_{ij}$ from 6 to 3. The assumption of plane magnetization reduces the number of independent components $M_{i}$ to 2 instead of 3. We denote by $\mathcal{MB} = \set{I_{k}}$ the initial integrity basis given in \autoref{tab:even-inv}. The set $\widetilde{\mathcal{MB}}$ of restrictions of polynomial functions $I_{k}$ to the subspace of $(\MM,\bsigma)$ which satisfy \eqref{eq:in-plane} spans a new algebra $\mathcal{A}$ of polynomial functions in 5 variables. Our goal is to reduce this set $\widetilde{\mathcal{MB}}$ of 30 invariants into a smaller generating set $\mathcal{G}=\set{\tilde{I_{l}}}$ of $\mathcal{A}$.

\section{Reduced sets of generators for the different fibers}
\label{sec:reduced-generating-sets}

Several algorithms/programs, able to check whether or not a set $\mathcal{B}=\set{I_{j}}$ of homogeneous polynomial invariants $I_{j}$ is an integrity basis, are available in the literature~\cite{Gor1900,Boe1987,DK2015,Oli2016,DOADK2020,TOD2021}. In the case of finite groups, these algorithms require the knowledge, \textit{a priori}, of a bound on the total degree of the generators. Moreover, these algorithms also allow for reducing an integrity basis $\mathcal{B}$ into a minimal integrity basis $\mathcal{MB}$, by checking linear relations among the invariants. In the present problem, a minimal integrity basis $\mathcal{MB}=\set{I_{k}}$ is already known~\cite{SSR1963,TOD2021} and recalled in \autoref{tab:even-inv}. Then, we evaluate all the $I_{k}$ for a given stress/magnetization state (here for $\MM\cdot \nn =0$ and $\bsigma\cdot \nn =0$). The restrictions of the function $I_{k}$ to this subspace, denoted by $\tilde{I}_{k}$, are however not linearly independent: they  satisfy some linear relations. An algorithm to compute these relations is provided in \autoref{sec:algo}, as well as its implementation in \emph{Macaulay2}~\cite{Macau2}, a software system devoted to computations in algebraic geometry and commutative algebra.

Finally, a minimal set of generators $\mathcal{G}=\set{\tilde{I}_{k_{l}}}$ of $\mathcal{A}$ can be produced, removing redundant generators from $\widetilde{\mathcal{MB}} = \set{\tilde{I}_{k}}$. This has been done for each in-plane magneto-mechanical loading \eqref{eq:plane:100}, \eqref{eq:plane:110} or \eqref{eq:plane:111}. Corresponding minimal sets of generators are denoted respectively by $\mathcal{G}_{\theta}$, $\mathcal{G}_{\alpha'}$, and $\mathcal{G}_{\gamma}$. The results are summarized in \autoref{tab:results}. The cardinalities of these sets are small (compared to the initial 30 invariants in \autoref{tab:even-inv}). In terms of modeling, this means that the Gibbs free energy density can be expressed, with no lack of generality, as a function of invariants $I_{k}$ such that $\tilde{I}_{k}$ belongs to $\mathcal{G}$ when in-plane stress and magnetization are considered:
\begin{equation*}
  \Psi = \Psi(I_{k_{1}}, \dotsc, I_{k_{L}}), \qquad \tilde{I}_{k_{l}} \in \mathcal{G}, \quad L = \text{card} \, \mathcal{G}.
\end{equation*}

\begin{rem}\label{rem:inclusion}
  Plane constitutive laws can be deduced either by restricting first the energy density to the plane state variables and then, by deriving them with respect to these variables to get the dual plane variables (a) or by selecting the in-plane terms of 3D constitutive laws restricted to a 2D loading (b). This is mathematically justified by the fact that the \emph{pullback} and the \emph{exterior derivative} commute~\cite[theorem 5.3]{Spi2018}.
\end{rem}

\begin{table}[h]
  \centering
  \begin{tabular}{ccL{7.5cm}} %
    \toprule
    Reduced basis           & Cardinal $L$ & List of $\octa$-invariants
    \\                                                                                                                                             \midrule
    $\mathcal{G}_{\theta}$  & 7            & $\tr \bsigma$, $I_{002}$, $I_{020}$, $I_{200}$ $I_{201}$, $I_{210}$, $I_{400}$                                                                                            \\
    \midrule
    $\mathcal{G}_{\alpha'}$ & 15           & $\tr \bsigma$, $I_{002}$, $I_{020}$, $I_{003}$, $I_{030}$, $I_{200}$, $I_{201}$, $I_{210}$, $I_{202}^a$, $I_{211}$, $I_{220}$, $I_{400}$, $I_{401}$, $I_{410}$, $I_{600}$
    \\
    \midrule
    $\mathcal{G}_{\gamma}$  & 8            & $\tr \bsigma$, $I_{020}$, $I_{030}$, $I_{200}$, $I_{210}$, $I_{220}$, $I_{410}$, $I_{600}$                                                                                \\
    \bottomrule
  \end{tabular}
  \caption{Reduced generating sets $\mathcal{G}=\set{\tilde{I}_{k_{l}}}$ for the different  textures.}
  \label{tab:results}
\end{table}

\subsection{\texorpdfstring{$\theta$}{theta}-fiber}
\label{subsec:100}

This is the most favorable texture orientation for a reduction of the integrity basis evaluated with in-plane conditions. The direction \{001\} is indeed the normal $\nn_{\theta}=\ee_3$ to the cubic crystal network (a symmetry plane of the microstructure). Because of the presence of null coefficients in \eqref{eq:plane:100} (especially on the off-diagonal part of $\bsigma$), 12 evaluated invariants in $\mathcal{MB}$ vanish, namely
\begin{equation*}
  I_{003} = I_{004} = I_{014} = I_{202}^b = I_{203} = I_{212}^b = I_{204} = I_{402} = I_{222} =I_{401}=I_{411}=I_{600}= 0.
\end{equation*}
In addition, we get the following 11 relations.
\begin{align*}
   & I_{012}=\frac{1}{6}I_{002}\tr \bsigma,
   &                                                                                                  & I_{030}=\frac{1}{18}\left( 9I_{020}\tr \bsigma-2(\tr \bsigma)^3\right),      \\
   & I_{022}=\frac{1}{18}\left(2I_{002}(\tr \bsigma)^2 -9I_{002}I_{020} \right),
   &                                                                                                  & I_{211}=\frac{1}{6}I_{201}\tr \bsigma,                                       \\
   & I_{212}^a =\frac{1}{6}\left(I_{002}I_{210}-2I_{202}^a\tr \bsigma \right) ,
   &                                                                                                  & I_{221}=\frac{1}{18} \left(2I_{201}(\tr \bsigma)^2 -9I_{020}I_{201} \right), \\
   & I_{213}=\frac{1}{36}\, I_{002}I_{201}\tr \bsigma ,
   &                                                                                                  & I_{202}^a = \frac{1}{6}\, I_{002}I_{200} ,                                   \\
   & I_{220}=\frac{1}{18} \left( 6I_{210}\tr \bsigma+3I_{020}I_{200}-2I_{200}(\tr \bsigma)^2 \right),
   &                                                                                                  & I_{410}=\frac{1}{6}  I_{400}\tr \bsigma  ,                                   \\
   & I_{601}=\frac{1}{18} \left( 9I_{201}I_{400} -4I_{200}^2I_{201} \right).
\end{align*}

\subsection{\texorpdfstring{$\alpha'$}{alpha'}-fiber}
\label{subsec:110}

$\nn_{\alpha'}$ is normal to another symmetry plane of the crystal network. This explains why the number of useful invariants is drastically reduced (this reduction is however smaller than for the $\theta$ fiber). For this configuration, we get the following 15 relations.
\begin{align*}
  {I_{012}}   & = \frac{1}{54} \left( -18I_{003}-18I_{030}+9I_{002}\tr \bsigma+9I_{020}\tr \bsigma-2(\tr \bsigma)^3 \right)                                                            \\
  {I_{004}}   & = \frac{1}{72}\left( 9(I_{002})^2+6I_{002}I_{020}-3(I_{020})^2 +12I_{003}\tr \bsigma \right.                                                                           \\
              & \quad \left. -12I_{012}\tr \bsigma+12I_{030}\tr \bsigma-2I_{002}(\tr \bsigma)^2-2I_{020}(\tr \bsigma)^2\right)
  \\
  {I_{022}}   & = \frac{1}{18} \left( -6I_{002}I_{020}+3(I_{020})^2 -12I_{030}\tr \bsigma+2I_{020}(\tr \bsigma)^2 \right)                                                              \\
  {I_{014}}   & = \frac{1}{18} \left( 6I_{020}I_{003}+9I_{002}I_{012}+3I_{020}I_{012}+6I_{002}I_{030} -2I_{002}I_{020}\tr \bsigma-2I_{012}(\tr \bsigma)^2\right)
  \\
  {I_{212}^a} & = \frac{1}{36} \left( 3I_{002}I_{210}-3I_{020}I_{210}+12I_{220}\tr \bsigma-2I_{210}(\tr \bsigma)^2\right)                                                              \\
  {I_{212}^b} & = \frac{1}{36}\left( -9I_{002}I_{201}+72I_{203}+18I_{221}-12I_{202}^b\tr \bsigma+12I_{211}\tr \bsigma+2I_{201}(\tr \bsigma)^2 \right)
  \\
  {I_{204}}   & = \frac{1}{18}\left( 6I_{003}I_{210}+9I_{002}I_{202}^a+3I_{020}I_{202}^a+6I_{002}I_{220}-2I_{002}I_{210}\tr \bsigma-2I_{202}^a(\tr \bsigma)^2 \right)
  \\
  {I_{213}}   & = \frac{1}{18}\left( 3I_{003}I_{201}+3I_{012}I_{201}-3I_{020}I_{202}^b -12I_{203}\tr \bsigma+2I_{202}^b(\tr \bsigma)^2\right)
  \\
  {I_{222}}   & = \frac{1}{18}\left( 3I_{002}I_{211} - 2I_{020}I_{201}\tr \bsigma +3I_{020}I_{211}  +3I_{003}I_{201}  +6I_{012}I_{201} \right.                                         \\
              & \quad \left.  +6I_{030}I_{201}  +2I_{202}^b(\tr \bsigma)^2  -2I_{211}(\tr \bsigma)^2 - 12I_{203}\tr \bsigma \right)
  \\
  {I_{202}^b} & = \frac{1}{18}\left( 3I_{002}I_{200}+3I_{020}I_{200}-18I_{202}^a -36I_{211}-18I_{220}+6I_{201}\tr \bsigma\right.                                                       \\
              & \quad \left.+6I_{210}\tr \bsigma-2I_{200}(\tr \bsigma)^2\right)                                                                                                        \\
  {I_{203}}   & = \frac{1}{72}\left( 12I_{003}I_{200}+24I_{012}I_{200}+12I_{030}I_{200}+9I_{002}I_{201}+3I_{020}I_{201}+3I_{002}I_{210}-3I_{020}I_{210} \right.
  \\
              & \quad \left. -4I_{002}I_{200}\tr \bsigma-4I_{020}I_{200}\tr \bsigma-12I_{202}^a\tr \bsigma-24I_{211}\tr \bsigma+2I_{201}(\tr \bsigma)^2+2I_{210}(\tr \bsigma)^2\right)
  \\
  {I_{221}}   & = \frac{1}{18}\left(-6I_{003}I_{200}-18I_{012}I_{200}-12I_{030}I_{200}-6I_{020}I_{201}+3I_{020}I_{210}+2I_{002}I_{200}\tr \bsigma\right.
  \\
              & \quad \left.+4I_{020}I_{200}\tr \bsigma +6I_{202}^a\tr \bsigma+6I_{202}^b\tr \bsigma+12I_{211}\tr \bsigma-2I_{201}(\tr \bsigma)^2-2I_{210}(\tr \bsigma)^2 \right)
  \\
  {I_{402}}   & = \frac{1}{72}\left( 9I_{201}^2+6I_{201}I_{210}-24I_{200}I_{211}+3I_{020}I_{400}+12I_{401}\tr \bsigma-2I_{400}(\tr \bsigma)^2  \right)
  \\
  {I_{411}}   & = \frac{1}{144}\left( -9I_{201}^2-6I_{201}I_{210}-24I_{200}I_{202}^b-24I_{200}I_{211}+12I_{002}I_{400}+9I_{020}I_{400}\right.                                          \\
              & \quad \left. +8I_{200}I_{201}\tr \bsigma+12I_{401}\tr \bsigma-6I_{400}(\tr \bsigma)^2 \right)                                                                          \\
  {I_{601}}   & = \frac{1}{18}\left( -4I_{200}^2I_{201}+6I_{201}I_{400}-3I_{210}I_{400}+6I_{200}I_{410}+I_{200}I_{400}\tr \bsigma-3I_{600}\tr \bsigma \right)
\end{align*}

\subsection{\texorpdfstring{$\gamma$}{gamma}-fiber}
\label{subsec:111}

In this configuration, stress tensor and magnetization do not have vanishing components (see~\eqref{eq:plane:111}). However, useful relations appear between the invariants involving the off-diagonal part of $\bsigma$. We get for this fiber the following 22 relations.

\begin{align*}
  {I_{002}}   & = \frac{1}{6}\left( 12I_{020}+(\tr \bsigma)^2  \right)                                                                                                      \\
  {I_{012}}   & = \frac{1}{12} \left(  -6I_{003}-I_{002}\tr \bsigma+9I_{020}\tr \bsigma \right)                                                                             \\
  {I_{003}}   & = \frac{1}{6}\left( 12I_{030}-I_{002}\tr \bsigma+5I_{020}\tr \bsigma \right)                                                                                \\
  {I_{004}}   & = \frac{1}{6} \left(  3I_{002}^2-18I_{002}I_{020}+27I_{020}^2+2I_{003}\tr \bsigma \right)                                                                   \\
  {I_{022}}   & = \frac{1}{6}\left( -2I_{002}^2+13I_{002}I_{020}-18I_{020}^2-2I_{003}\tr \bsigma \right)                                                                    \\
  {I_{202}^a} & = \frac{1}{3} \left( -3I_{220}+I_{210}\tr \bsigma \right)                                                                                                   \\
  {I_{212}^a} & = \frac{1}{6} \left( 4I_{002}I_{210}-9I_{020}I_{210}-2I_{202}^a\tr \bsigma \right)                                                                          \\
  {I_{212}^b} & = \frac{1}{12}\left( -3I_{002}I_{201}+3I_{020}I_{201}+12I_{203} -I_{202}^b\tr \bsigma+6I_{211}\tr \bsigma \right)                                           \\
  {I_{203}}   & = \frac{1}{12}\left( +2I_{002}I_{201}-I_{020}I_{201}-6I_{221}-4I_{211}\tr \bsigma \right)                                                                   \\
  {I_{204}}   & = \frac{1}{6}\left(  2I_{003}I_{210}-6I_{002}I_{202}^a+18I_{020}I_{202}^a+3I_{002}I_{210}\tr \bsigma-9I_{020}I_{210}\tr \bsigma \right)
  \\
  {I_{213}}   & = \frac{1}{24} \left( 2I_{003}I_{201}-32I_{002}I_{211}+72I_{020}I_{211} +3I_{002}I_{201}\tr \bsigma -3I_{020}I_{201}\tr \bsigma-16I_{203}\tr \bsigma\right)
  \\
  {I_{402}}   & = \frac{1}{36} \left( 4I_{020}I_{200}^2+9I_{201}^2-12I_{210}^2-12I_{200}I_{202}^a-3I_{002}I_{400} \right.
  \\
              & \quad \left. -6I_{020}I_{400}+4I_{200}I_{210}\tr \bsigma-12I_{410}\tr \bsigma \right)
  \\
  {I_{222}}   & = \frac{1}{54} \left(  8I_{002}I_{201}\tr \bsigma  +9I_{002}I_{202}^b- 108I_{002}I_{211}- 45I_{020}I_{202}^b+ 216I_{020}I_{211} \right.
  \\
              & \quad \left. +3I_{003}I_{201}- 12I_{012}I_{201} - 36I_{203}\tr \bsigma \right)
  \\
  {I_{014}}   & = \frac{1}{12}\left( 6I_{002}I_{003}-14I_{020}I_{003} +I_{002}^2\tr \bsigma-6I_{002}I_{020}\tr \bsigma+9I_{020}^2\tr \bsigma \right)
  \\
  {I_{201}}   & = \frac{1}{6}\left( 12I_{210}+I_{200}\tr \bsigma\right)                                                                                                     \\
  {I_{202}^b} & = \frac{1}{6}\left( -4I_{002}I_{200}+9I_{020}I_{200}-12I_{202}^a+3I_{201}\tr(\bsigma \right)                                                                \\
  {I_{211}}   & = \frac{1}{12}\left( I_{002}I_{200}+12I_{202}^a -I_{201}\tr \bsigma\right)                                                                                  \\
  {I_{221}}   & = \frac{1}{36}\left( -6I_{003}I_{200}-12I_{002}I_{201}+24I_{020}I_{201}+I_{002}I_{200}\tr \bsigma+12I_{202}^a\tr \bsigma  \right)
  \\
  {I_{400}}   & = \frac{1}{2}I_{200}^2                                                                                                                                      \\
  {I_{401}}   & = \frac{1}{12}\left( 6I_{200}I_{201} -24I_{410}-I_{200}^2\tr \bsigma \right)                                                                                \\
  {I_{411}}   & = \frac{1}{72}\left( I_{002}I_{200}^2-6I_{201}^2+12I_{200}I_{202}^a+6I_{401}\tr \bsigma \right)                                                             \\
  {I_{601}}   & = \frac{1}{36}\left( I_{200}^2I_{201}-6I_{200}I_{401} -6I_{600}\tr \bsigma\right)
\end{align*}

\subsection{Combination of \texorpdfstring{$\theta$}{theta}, \texorpdfstring{$\alpha'$}{alpha'} and \texorpdfstring{$\gamma$}{gamma} fibers}
\label{subsec:mix}

As explained in introduction of this work, it is usually possible to find a set of fibers that describes the texture of a material. If the three previous fibers are combined to form the texture of a thin layer material subjected to an in-plane magneto-mechanical loading, the invariant generators in this situation is the union of the invariant generators of each fiber. Then it can be noticed that the list of $\octa$-invariants of the reduced generating sets of $\theta$ and $\gamma$ fibers are included in the list of $\octa$-invariants of the reduced generating set of the $\alpha'$ fiber. This means that the list of 15 invariants associated with the reduced generating set $\mathcal{G}_{\alpha'}$ is the reduced generating set of any combination of the three preceding fibers.

\section{Conclusion}

The magneto-elasticity of cubic ferromagnetic materials is described using a Gibbs free energy density $\Psi = \Psi(I_{k})$
defined as a function of well-chosen cubic invariants $I_{k}$ of the stress $\bsigma$ and the magnetization $\MM$. It is relevant to take these cubic invariants into account for every crystallographic texture/fiber, with the drawback that the corresponding minimal integrity basis is constituted of a quite large number (30) of invariants~\cite{SSR1963,TOD2021}.

The magneto-mechanical coupled behavior of textured (cubic) ferromagnetic materials subjected to in-plane magneto-mechanical loadings has been streamlined. We have shown that for such loadings, and for specific textures/fibers, the Gibbs free energy density can be written as a function of a lower number of invariants
\begin{equation*}
  \Psi = \Psi(I_{k_1}, \dotsc, I_{k_L}),
  \qquad
  L<30.
\end{equation*}
Indeed, we have computed relations among the fundamental cubic invariants $I_{k}$ which arise when they are restricted to plane magneto-elasticity problems. We have obtained, this way, some reduced (minimal) sets of cardinal $L$ of cubic generators $I_{k_l}$ for three major fibers, namely $\theta$ ($L=7$), $\alpha'$ ($L=15$) and $\gamma$ ($L=8$) and their combination ($L=15$). To do so, we have adapted an algorithm initially proposed to prove the minimality of an integrity basis in~\cite{TOD2021}. An implementation of this algorithm in Macaulay2, a software devoted to algebraic geometry and commutative algebra~\cite{Macau2}, has also been provided.

\appendix

\section{Reduction algorithm and implementation}
\label{sec:algo}

\subsection{Algorithm}

The algorithm we propose here generates all the relations between the homogeneous polynomials $\set{\tilde{I}_{k}}$ in $\widetilde{\mathcal{MB}}$, the set of cubic invariants $\set{I_{k}}$ restricted to plane loadings (plane defined by the normal unit vector $\nn$). For instance, for the fiber $\theta$ ($\nn_\theta = \ee_{3}$), $M_3=0$ and $\sigma_{13}=\sigma_{23}=\sigma_{33}=0$. If we choose $\widetilde{\ee}_{1}=\ee_{1}$, $\widetilde{\ee}_2=\ee_2$ as a natural basis of the subspace $M_3=0$ of vectors $\MM$ and
\begin{equation*}
  \widetilde{\be}_{\text{I}}=\text{\ee}_{11}=
  \begin{pmatrix}
    1 & 0 & 0 \\
    0 & 0 & 0 \\
    0 & 0 & 0
  \end{pmatrix}, \qquad
  \widetilde{\be}_{\text{II}}=\text{\ee}_{22}=
  \begin{pmatrix}
    0 & 0 & 0 \\
    0 & 1 & 0 \\
    0 & 0 & 0
  \end{pmatrix}, \qquad
  \widetilde{\be}_{\text{III}}=\text{\ee}_{12}=
  \begin{pmatrix}
    0 & 1 & 0 \\
    1 & 0 & 0 \\
    0 & 0 & 0
  \end{pmatrix}
\end{equation*}
as a natural basis of the subspace $\bsigma.\ee_3=0$ of stresses $\bsigma$, we get
\begin{equation*}
  \widetilde{\MM} = \widetilde{M}_{1}\widetilde{\ee}_{1}+\widetilde{M}_2\widetilde{\ee}_2 \quad \text{and} \quad \widetilde{\bsigma} = \widetilde{\sigma}_1\widetilde{\be}_{\text{I}}+\widetilde{\sigma}_{2}\widetilde{\be}_{\text{II}}+\widetilde{\sigma}_{3}\widetilde{\be}_{\text{III}}.
\end{equation*}
For each considered fiber, the restricted invariants are expressed as polynomial functions of $\widetilde{\bsigma}$ and $\widetilde{\MM}$ as above. These polynomial functions are homogeneous, both in $\widetilde{\bsigma}$ and $\widetilde{\MM}$. For such a \emph{bi-homogeneous polynomial}, we introduce the degree $\alpha = \textrm{deg}(\widetilde{\MM})$ in $\widetilde{\MM}$ and the degree $\beta = \textrm{deg}(\widetilde{\bsigma})$ in $\widetilde{\bsigma}$. This bi-degree is denoted by $(\alpha, \beta)$ and the total degree by $d=\alpha+\beta$. A bi-homogeneous polynomial of bi-degree $(\alpha, \beta)$ is thus a linear combination of monomials
\begin{equation}\label{eq:monomial}
  m_{\balpha,\bbeta}=\widetilde{M}_{1}^{\alpha_{1}}\widetilde{M}_{2}^{\alpha_2}\widetilde{\sigma}_{1}^{\beta_{1}}\widetilde{\sigma}_{2}^{\beta_2}\widetilde{\sigma}_{3}^{\beta_3}
\end{equation}
with $\balpha=(\alpha_{1}, \alpha_2)$ ($\alpha_{k}$ being the degree in $\widetilde{M}_{k}$, so that $\alpha=\abs{\balpha}=\alpha_{1}+\alpha_2$) and $\bbeta=(\beta_{1}, \beta_{2}, \beta_{3})$ ($\beta_{l}$ being the degree in $\widetilde{\sigma}_{l}$, so that $\beta=\abs{\bbeta}=\beta_{1}+\beta_2+\beta_3$).

On the set of bi-degrees, we introduce the \emph{degree lexicographic order} as follows
\begin{multline*}
  (\alpha, \beta) < (\alpha^\prime, \beta^\prime) \quad \text{if} \quad \alpha+\beta < \alpha^\prime+\beta^\prime
  \\
  \text{or} \quad \alpha+\beta = \alpha^\prime+\beta^\prime \quad \text{and} \quad  (\alpha < \alpha^\prime \quad \text{or} \quad  \alpha = \alpha^\prime \quad \text{and} \quad \beta < \beta^\prime).
\end{multline*}
Using this order relation, we define $(\alpha, \beta)_{1}$ and $(\alpha,\beta)_{J_{Max}}$ respectively, as the least and the greatest bi-degree $(\alpha, \beta)$ appearing in the finite list of evaluated bi-homogeneous polynomials $\widetilde{\mathcal{MB}}$, where it is understood that each evaluated polynomial which vanishes is removed from $\widetilde{\mathcal{MB}}$. We can therefore partition $\widetilde{\mathcal{MB}}$ into bi-graded sets
\begin{equation*}
  \widetilde{\mathcal{MB}} = \widetilde{\mathcal{MB}}_{(\alpha, \beta)_{1}} \cup \dotsb \cup \widetilde{\mathcal{MB}}_{(\alpha, \beta)_{J_\textrm{Max}}},
\end{equation*}
using the following partition of the integrity basis $\mathcal{MB}$ in \autoref{tab:even-inv}, and where the set ${\mathcal{MB}}_{\alpha, \beta}$ consists in bi-homogeneous polynomials in $({\bsigma},{\MM})$ of bi-degree $(\alpha, \beta)$.
\begin{align*}
   & \mathcal{MB}_{0,1}=\set{\tr\bsigma},              &  & \mathcal{MB}_{0,2}=\set{I_{020},I_{002}},
  \\
   & \mathcal{MB}_{0,3}=\set{I_{012},I_{003},I_{030}}, &  & \mathcal{MB}_{0,4}=\set{I_{004},I_{022}},
  \\
   & \mathcal{MB}_{0,5}=\set{I_{014}},                 &  & \mathcal{MB}_{2,0}=\set{I_{200}},
  \\
   & \mathcal{MB}_{2,1}=\set{I_{201},I_{210}},         &  & \mathcal{MB}_{2,2}=\set{I_{202}^a,I_{202}^b,I_{211},I_{220}},
  \\
   & \mathcal{MB}_{2,3}=\set{I_{210},I_{201}},         &  & \mathcal{MB}_{2,4}=\set{I_{203},I_{212}^a,I_{212}^b,I_{221}},
  \\
   & \mathcal{MB}_{4,0}=\set{I_{400}},                 &  & \mathcal{MB}_{4,1}=\set{I_{401},I_{410}},
  \\
   & \mathcal{MB}_{4,2}=\set{I_{402},I_{411}},         &  & \mathcal{MB}_{6,0}=\set{I_{600}},
  \\
   & \mathcal{MB}_{6,1}=\set{I_{601}}.
\end{align*}

The set of all bi-homogeneous polynomials in $(\widetilde{\MM},\widetilde{\bsigma})$  having the same bi-degree $(\alpha, \beta)$ is a finite dimensional vector space and the subspace of such polynomials which belong to the algebra $\mathcal{A}$ is denoted by $\mathcal{A}_{\alpha, \beta}$. Hence $\mathcal{A}$ can be written as the direct sum
\begin{equation*}
  \mathcal{A} = \bigoplus_{\alpha, \beta} \mathcal{A}_{\alpha, \beta}.
\end{equation*}
Each finite dimensional vector space $\mathcal{A}_{\alpha, \beta}$ is spanned (as a vector space) by $\widetilde{\mathcal{MB}}_{\alpha, \beta}$ (which is empty if $(\alpha, \beta) > (\alpha,\beta)_{J_{Max}}$) and the so-called \emph{reducible elements} of bi-degree $(\alpha, \beta)$ which can be written as products of (at least two) elements in
\begin{equation*}
  \bigcup_{(\mu,\nu) < (\alpha,\beta)} \widetilde{\mathcal{MB}}_{(\mu,\nu)}.
\end{equation*}
Hence, a minimal set of generators $\mathcal{G}$ of $\mathcal{A}$ can be extracted from $\widetilde{\mathcal{MB}}$ by choosing a vectorial basis in each $\mathcal{A}_{(\alpha, \beta)_{J}}$ ($1 \le J \le J_{Max}$) and by eliminating \emph{arbitrarily} some generators $\widetilde{I_{k}}$. This can be done, bi-degree by bi-degree $(\alpha,\beta)$, once one knows exactly the linear relations between the $\widetilde{I_{k}}$ and the reducible elements of bi-degree $(\alpha,\beta)$.

To do so, we proceed recursively as follows. Given $1 \le J \le J_{Max}$, let $\left(P_{1}, \dotsc ,P_{N}\right)$ be the $N$ elements $P_{j}$ in $\mathcal{R}_{J} \cup \widetilde{\mathcal{MB}}_{(\alpha, \beta)_{J}}$, where $\mathcal{R}_{J}$ is the set of reducible elements of bi-degree $(\alpha,\beta)_{J}$. Then, let $m_{1}, \dotsc ,m_{M}$ be the monomials which appear in the family of bi-homogeneous polynomials $\left(P_{1}, \dotsc ,P_{N}\right)$. Each of these monomials is written $m_{(\balpha,\bbeta)}$ with $\abs{\balpha}=\alpha$ and $\abs{\bbeta}=\beta$ (see \eqref{eq:monomial}). We can thus write
\begin{equation*}
  P_{j} = \sum_{i} A_{ij} m_{i},
\end{equation*}
where $A := (A_{ij})$ is an $M \times N$ matrix. The computation of a basis $\uu^{k} := \sum_{i} u^{k}_{i} m_{i}$, where $1 \le k \le N-R$ and $R : = \mathrm{rank}(A)$, furnishes all the independent linear relations between the $P_{j}$, which are written
\begin{equation}\label{eq:Mu0}
  \sum_{j} u^{k}_{j} P_{j}  = 0, \qquad 1 \le k \le N-R.
\end{equation}
Thanks to the knowledge of these relations, we are able to extract, \emph{by hand}, a minimal set
\begin{equation*}
  \mathcal{G}_{(\alpha, \beta)_{J}} \subset \widetilde{\mathcal{MB}}_{(\alpha, \beta)_{J}},
\end{equation*}
which spans (together with $\mathcal{R}_{J}$) the vector space $\mathcal{A}_{(\alpha, \beta)_{J}}$. The union
\begin{equation*}
  \mathcal{G}=\mathcal{G}_{(\alpha, \beta)_{1}} \cup \dotsb \cup \mathcal{G}_{(\alpha, \beta)_{J}} \cup \dotsb \cup \mathcal{G}_{(\alpha, \beta)_{J_\textrm{Max}}} ,
\end{equation*}
is then the sought minimal set of generators of the algebra $\mathcal{A}$.

For a given in-plane loading defined by a normal $\nn$, and given by~\eqref{eq:in-plane}, the following algorithm produces the expected finite list of relations $\mathcal{L}$ among the set of restricted invariants $\widetilde{\mathcal{MB}} = \set{\tilde{I}_{k}}$.

\begin{itemize}
  \item \textbf{Input:} The set $\widetilde{\mathcal{MB}} = \set{\tilde{I}_{k}}$ of restricted invariants.

  \item \textbf{Output:} A list $\mathcal{L}_{J_\textrm{Max}}$ of polynomials relations between these restricted invariants $\tilde{I}_{k}$.

  \item \textbf{Initialization:} $J=0$,  $\mathcal{L}_0:=\emptyset$.

  \item \textbf{For $ 1 \le J \le J_{Max}$:}
        \begin{enumerate}
          \item Generate the family $\mathcal{R}_{J}$ of all \emph{reducible} homogeneous polynomials of bi-degree $(\alpha, \beta)_{J}$;
          \item Compute a basis $(\uu^{k})$ ($1 \le k \le N-R$) of the kernel of the matrix $A$;
          \item Update the list $\mathcal{L}_{J}:=\left[\mathcal{L}_{J-1},[   \sum u_{j}^1P_{j} ,\dotsc, \sum u_{j}^{N-R}P_{j} ] \right]$;
        \end{enumerate}
  \item \textbf{Return} $\mathcal{L}_{J_{Max}}$.
\end{itemize}

\subsection{Implementation in Macaulay2}

The code presented here is in the Macaulay2 language (see~\cite{Macau2}) and can be run using a friendly web interface of Macaulay2 at
\begin{center}
  \url{https://www.unimelb-macaulay2.cloud.edu.au/#home}.
\end{center}
The invariants in \autoref{tab:even-inv} are computed for a particular form of $\bsigma$ and $\MM$. This is the \textbf{Input} step of the algorithm detailed above. The bounds $\mathrm{dMax} = 7$ and $\alpha_{\mathrm{Max}} = 6$ correspond respectively to the highest total degree and the highest partial degree in magnetization in the list $\mathcal{MB}$ of invariants $I_{k}$. The increment $\mathrm{k}$ is associated to the total degree and $\alpha$ to the degree in $\tilde{\MM}$.

\bigskip
\footnotesize

\begin{verbatim}
-------------------------------------Input--------------------------------------
--definition of the algebra Alg of evaluated polynomials
Alg = QQ[sig1, sig2, sig3, m1, m2, Degrees=>{{0,1},{0,1},{0,1},{1,0},{1,0}}]
--off-diagonal part
dbar=(b)->(matrix{{0,b_(0,1),b_(0,2)},{b_(1,0),0,b_(1,2)},{b_(2,0),b_(2,1),0}});
--deviatoric diagonal part
ddev = (b)->(matrix{{b_(0,0) - 1/3*trace(b), 0, 0},{0, b_(1,1) -
  1/3*trace(b), 0},{0, 0, b_(2,2) - 1/3*trace(b)}}) ;
--magnetization and stress bases for fiber theta
e1=matrix({{1_Alg},{0},{0}}); e2=matrix({{0},{1_Alg},{0}})
eI=matrix{{1_Alg,0,0},{0,0,0},{0,0,0}}
eII=matrix{{0,0,0},{0,1_Alg,0},{0,0,0}}
eIII=matrix{{0,1_Alg,0},{1_Alg,0,0},{0,0,0}}
--vectors and matrixes are omitted for others textures
--stress and magnetization for a given fiber
M = m1*e1+m2*e2
sig = sig1*eI+sig2*eII+sig3*eIII ; sigd = ddev(sig) ; sigdbar = dbar(sig)
--evaluation of the invariants
I010 = trace(sig)
I002=trace(sigdbar^2)
I020=trace(sigd*sigd)
I003=trace(sigdbar*sigdbar*sigdbar)
I012=trace(sigdbar^2*sigd)
I030=trace(sigd*sigd*sigd)
I004=trace(dbar(sigdbar^2)*dbar(sigdbar^2))
I022=trace(sigdbar*sigd*sigdbar*sigd)
I014=trace(sigdbar*dbar(sigdbar^2)*sigdbar*sigd)
I200=trace(transpose(M)*M)
I201=trace(dbar(M*transpose(M))*sigdbar)
I210=trace(ddev(M*transpose(M))*sigd)
I202a=trace(ddev(M*transpose(M))*(sigdbar^2))
I202b=trace(dbar(M*transpose(M))*dbar(sigdbar^2))
I211=trace(dbar(M*transpose(M))*sigdbar*sigd)
I220=trace((ddev(M*transpose(M)))*(sigd^2))
I203=trace(dbar(M*transpose(M))*dbar(sigdbar^2)*sigdbar)
I212a=trace(ddev(M*transpose(M))*ddev(sigdbar^2)*sigd)
I212b=trace(dbar(M*transpose(M))*dbar(sigdbar^2)*sigd)
I221=trace(dbar(M*transpose(M))*sigd*sigdbar*sigd)
I204=trace((ddev(M*transpose(M)))*sigdbar*dbar(sigdbar^2)*sigdbar)
I213=trace(dbar(M*transpose(M))*ddev(sigdbar^2)*sigdbar*sigd)
I400=trace(dbar(M*transpose(M))*dbar(M*transpose(M)))
I401=trace(dbar(M*transpose(M))*sigdbar*dbar(M*transpose(M)))
I410=trace(dbar(M*transpose(M))*sigd*dbar(M*transpose(M)))
I402=trace(dbar(M*transpose(M))*dbar(sigdbar^2)*dbar(M*transpose(M)))
I411=trace(dbar(M*transpose(M))*sigd*sigdbar*dbar(M*transpose(M)))
I600=trace(dbar(M*transpose(M))*dbar(M*transpose(M))*dbar(M*transpose(M)))
I601 = trace(ddev(M*transpose(M))*dbar(M*transpose(M))*ddev(M*transpose(M))*sigdbar)
I222=trace(dbar(M*transpose(M))*sigd*dbar(sigdbar^2)*sigd)
--creation of the list of evaluated invariants
MBtilde=new HashTable from {"I010"=>I010,"I002"=>I002,"I020"=>I020,"I003"=>I003,
	"I012"=>I012,"I030"=>I030,"I004"=>I004,"I022"=>I022,"I014"=>I014,
	"I200"=>I200,"I201"=>I201,"I210"=>I210,"I202a"=>I202a,"I202b"=>I202b,
	"I211"=>I211,"I220"=>I220,"I203"=>I203,"I212a"=>I212a,"I212b"=>I212b,
	"I221"=>I221,"I204"=>I204,"I213"=>I213,"I400"=>I400,"I401"=>I401,
	"I410"=>I410,"I402"=>I402,"I411"=>I411,"I600"=>I600,"I601"=>I601,
	"I222"=>I222}
--removal of vanishing invariants
MBtilde=delete(0_Alg,MBtilde)
--list of bi-degrees of MBtilde
MBtildeValues=values(MBtilde)
listDeg =apply(MBtildeValues,degree)
MBtildeIndex=keys(MBtilde)
--------------------------------------------------------------------------------
bound={7,6}
dMax=bound_0
alphaMax=bound_1
--definition of the free algebra FreeAlg
FreeAlg=QQ[MBtildeIndex,Degrees=>listDeg]
LJ={}
for k in 1..dMax do (
    for alpha in 0..alphaMax do (
        if k-alpha >= 0 then (
------------------------------------Step 1--------------------------------------
            Base=toString(basis({alpha,k-alpha},FreeAlg)),
            Pj=Base,
------------------------------------Step 2--------------------------------------
            for i in 0..length(MBtildeIndex)-1 do (
                Pj=replace(toString(MBtildeIndex_i),concatenate("
                (",(toString(MBtildeValues_i)),")"),Pj)
            ),
            Pj=value Pj,
            if Pj!=0_Alg then (
                mi=monomials(Pj),
                (m,A)=coefficients(Pj,Monomials=>mi),
                uj=generators(ker A),
------------------------------------Step 3--------------------------------------
                ujPj=(value(Base))*value(toString(uj)),
                if ujPj!=0 then (
                    LJ=append(LJ,ujPj)) ,
            )		
        )
    )
)
"LJMax"<<toString(LJ)<<close
\end{verbatim}


\end{document}